\documentclass[twoside,11pt,preprint]{article}
\usepackage{blindtext}
\usepackage{graphicx} 
\usepackage{amssymb}
\usepackage{amsmath, nccmath}
\usepackage{amsthm}
\usepackage{jmlr2e}
\usepackage{xcolor}
\usepackage{enumitem}
\usepackage[autostyle]{csquotes}

\usepackage{algorithm}
\usepackage{algpseudocode} 
\usepackage{siunitx}
\usepackage{ulem} 

\usepackage[outline]{contour} 
\contourlength{1.4pt}
\usepackage{bm}
\usepackage{hyperref}
\usepackage{hyperxmp}

\hypersetup{
  pdfauthor={Kenneth Bogert, Matthew Kothe},
  pdftitle={The Principle of Uncertain Maximum Entropy},
  pdfsubject={The Principle of Maximum Entropy is a rigorous technique for estimating an unknown distribution given partial information while simultaneously minimizing bias.  However, an important requirement for applying the principle is that the available information be provided error-free.  We relax this requirement using a memoryless communication channel as a framework to derive a new, more general principle.  We show our new principle provides an upper bound on the entropy of the unknown distribution and the amount of information lost due to the use of a given communications channel is unknown unless the unknown distribution's entropy is also known. Using our new principle we provide a new interpretation of the classic principle and experimentally show its performance relative to the classic principle and some other generally applicable solutions.
},
  pdfkeywords={maximum entropy, uncertainty, information theory, information geometry}
}

\usepackage{totpages}
\jmlrheading{26}{2026}{1-\pageref{TotPages}}{1/26}{1/26}{21-0000}{Kenneth Bogert and Matthew Kothe}

\ShortHeadings{The Principle of Uncertain Maximum Entropy}{Bogert and Kothe}
\firstpageno{1}

\newcommand{\ols}[1]{\mskip.5\thinmuskip\overline{\mskip-.5\thinmuskip {#1} \mskip-.5\thinmuskip}\mskip.5\thinmuskip} 
\newcommand{\olsi}[1]{\,\overline{\!{#1}}} 
\makeatletter
\newcommand\closure[1]{
  \tctestifnum{\count@stringtoks{#1}>1} 
  {\ols{#1}} 
  {\olsi{#1}} 
}
\long\def\count@stringtoks#1{\tc@earg\count@toks{\string#1}}
\long\def\count@toks#1{\the\numexpr-1\count@@toks#1.\tc@endcnt}
\long\def\count@@toks#1#2\tc@endcnt{+1\tc@ifempty{#2}{\relax}{\count@@toks#2\tc@endcnt}}
\def\tc@ifempty#1{\tc@testxifx{\expandafter\relax\detokenize{#1}\relax}}
\long\def\tc@earg#1#2{\expandafter#1\expandafter{#2}}
\long\def\tctestifnum#1{\tctestifcon{\ifnum#1\relax}}
\long\def\tctestifcon#1{#1\expandafter\tc@exfirst\else\expandafter\tc@exsecond\fi}
\long\def\tc@testxifx{\tc@earg\tctestifx}
\long\def\tctestifx#1{\tctestifcon{\ifx#1}}
\long\def\tc@exfirst#1#2{#1}
\long\def\tc@exsecond#1#2{#2}
\makeatother

\newcommand{\W}[0]{\mathcal{W}}
\newcommand{\X}[0]{\mathcal{X}}
\newcommand{\Y}[0]{\mathcal{Y}}
\newcommand{\Z}[0]{\mathcal{Z}}

\newcommand{\truePW}[0]{P_0(W)}
\newcommand{\empPW}[0]{\tilde{P}(W)}
\newcommand{\barPW}[0]{\closure{P}(W)}

\newcommand{\lambdaphiPW}[0]{P_{\lambda,\phi}(W)}
\newcommand{\dotphiPW}[0]{P_{\centerdot ,\phi}}
\newcommand{\PW}[0]{P(W)}

\newcommand{\truePX}[0]{P_0(X)}

\newcommand{\truePY}[0]{P_0(Y)}
\newcommand{\empPY}[0]{\tilde{P}(Y)}

\newcommand{\PY}[0]{P(Y)}

\newcommand{\PYX}[0]{P(Y | X)} 
\newcommand{\PXY}[0]{P(X | Y)}

\newcommand{\PXW}[0]{P(X | W)} 
\newcommand{\PWX}[0]{P(W | X)}

\newcommand{\PYW}[0]{P(Y | W)} 

\newcommand{\truePrw}[0]{Pr_0(w)}
\newcommand{\empPrw}[0]{\tilde{Pr}(w)}
\newcommand{\barPrw}[0]{\closure{Pr}(w)}

\newcommand{\lambdaphiPrw}[0]{Pr_{\lambda,\phi}(w)}

\newcommand{\Prw}[0]{Pr(w)}

\newcommand{\barPry}[0]{\closure{Pr}(y)}
\newcommand{\Pry}[0]{Pr(y)}

\newcommand{\Prxw}[0]{Pr(x | w)}

\newcommand{\Pryx}[0]{Pr(y | x)}

\newcommand{\feat}[0]{\phi_k(w)}

\newcommand{\algname}[0]{\texttt}

\newcommand{\argmax}{\mathop{\mathrm{argmax}}\limits}
\newtheorem{kdb_theorem}{Theorem}
\newtheorem{kdb_corollary}{Corollary}[kdb_theorem]

\begin{document}

\title{The Principle of Uncertain Maximum Entropy}

\author{\name Kenneth Bogert \email kbogert@unca.edu 
       \AND
       \name Matthew Kothe \email mkothe@unca.edu \\
       \addr Department of Computer Science\\
       University of North Carolina Asheville\\
       1 University Heights, Asheville, NC 28801, USA}

\editor{My editor}

\maketitle

\begin{abstract}
The Principle of Maximum Entropy is a rigorous technique for estimating an unknown distribution given partial information while simultaneously minimizing bias.  However, an important requirement for applying the principle is that the available information be provided error-free~\citep{jaynes_rationale_1982}.  We relax this requirement using a memoryless communication channel as a framework to derive a new, more general principle.  We show our new principle provides an upper bound on the entropy of the unknown distribution and the amount of information lost due to the use of a given communications channel is unknown unless the unknown distribution's entropy is also known. Using our new principle we provide a new interpretation of the classic principle and experimentally show its performance relative to the classic principle and some other generally applicable solutions.
\end{abstract}


\section{Introduction}

We consider the problem of estimating an unknown probability distribution given samples taken from it that, due to noise or other mechanisms, do not provide a consistent estimation of the original distribution.  We use a common memoryless communication channel model to ground our problem, in which a sender encodes a message sampled from the unknown distribution and transmits it over a possibly noisy channel to a receiver.  However, we may have additional information about the unknown distribution in addition to the received samples that we also wish to incorporate into the estimate.  Specifically, the structure of the unknown distribution may be known up to a set of parameters, reducing the problem to estimating the parameters of a so-structured distribution given the samples.  To this end, we wish to employ the Principle of Maximum Entropy to find parameters producing a satisfying distribution with the least bias possible.

The Principle of Maximum Entropy states that among all alternative distributions that equally match available information the best choice is the one with the maximum entropy.  As formalized by \cite{jaynes1957information} and \cite{shannon} and commonly encountered in practice in various disciplines the information takes the form of constraints on expectations over feature functions of the distribution variable.  The feature functions determine the structure of the estimated distribution and this configuration underpins a number of well known distributions that have the maximum entropy given their parameters, including the Gaussian and Gamma~\citep{cover2006elements}.

However, an important assumption for the application of the principle to real-world scenarios is that the empirical feature expectations being constrained to are known~\citep{jaynes_rationale_1982}.  Or more restrictively requiring that there is only one possible empirical distribution that matches given information.  But, this requirement may be impossible to satisfy in domains where significant noise or otherwise uncertainty exists surrounding the samples, though special cases do exist~\citep{jaynes_rationale_1982}.  Previous attempts to address this problem assume the uncertainty is due to hidden variables~\citep{wang2012,bogert2022,bogert2016,hunsop_hong_maximum-entropy_2008} and use an expectation derived from the available error-free information to fill in the missing portion. We instead seek a more generally applicable principle that retains the desirable properties of the classic principle while uncertainty exists as to the empirical distribution as a result of the communication channel in use. 

Our contributions are as follows,

\begin{itemize}
  \setlength{\itemsep}{0em}
  \item We derive our new principle starting from an application of the classic principle to a noisy communication channel
  \item Through analysis we show our principle generalizes earlier principles and interpret the classic principle in light of our new one
  \item We present our new principle as a bilevel program and provide a single level solution
  \item We show our new principle produces bounds on the entropy of the unknown distribution and the amount of information lost when a given noisy communication channel is in use
  \item We present a number of experimental results in randomly generated problems that demonstrate the performance of our new principle as well as a relaxation useful when the provided samples are limited
\end{itemize}

\section{Background}

In this work, we will use the following conventions and notation.  Random variables are indicated with an uppercase letter such as $W$ with the corresponding lowercase letter $w$ giving an individual state the variable may be in from a set of all possible values indicated in script $\W$.  A probability distribution for $W$ is given by $\PW$, with each probability $1 \geq Pr(W = w) = Pr(w) \geq 0$.  We use a subscript to indicate parameters of a distribution $\lambdaphiPW$ with \enquote{ground truth} distributions indicated with a 0 ($\truePW , \truePrw$).  To distinguish two distributions over the same random variable we may mark one with an overline $\barPW$.  Empirical distributions are indicated with a tilde diacritic $\empPrw$, and may be defined in terms of sampling; for example, suppose that we produce $N$ i.i.d. samples from $\truePW$ and collect them into a multiset $\mathcal{S}$.  We may compute the resulting empirical distribution as $\empPrw \triangleq {\frac{1}{N}}\sum\nolimits_{w' \in \mathcal{S}} \delta (w, w') ~~\forall~ w \in \W$, where $\delta$ is the Kronecker delta. Finally, algorithm names will be denoted with \algname{TeleType}.

\subsection{Principle of Maximum Entropy}
Suppose we wish to estimate an unknown $\PW$ given knowledge of its moments or expectations.  To this end, we define $K$ feature functions $\phi_k \colon \W \to \mathbb{R}$ that represent the relevant information we have about each $w$.  We assume that the set of $\phi$ is correct and complete. We now require that our estimated distribution is constrained to match the expected features:  
\begin{equation}
\sum\limits_{w \in \W} \Prw \feat = \sum\limits_{w \in \W} \barPrw \feat
  \label{eqn:featureexpectations}
\end{equation}
or $\Phi = \closure{\Phi}$ for short, where $\barPrw$ is any distribution in the linear set that all equally produce the given $\closure{\Phi}$.  Unfortunately, however, there is a serious issue with this approach:  the distribution $\PW$ may be under-constrained, possibly leading to an infinite set of feasible $\PW$ for any given $\closure{\Phi}$.    To resolve this the Principle of Maximum Entropy states that we should choose the one distribution among the feasible set with the maximum entropy, ensuring it contains the least information of all alternatives and thus is expected to be the least biased.  The Principle of Maximum Entropy with feature expectation constraints is expressed as the following convex non-linear program:

\begin{equation}
\begin{array}{l}
 \max -\sum\nolimits_{w \in \W} \Prw \log \Prw \\
  \mbox{{\bf subject to}}\\
  \sum \nolimits_{w \in \W} \Prw = 1 \\
  \sum \nolimits_{w \in \W} \Prw \feat  = \sum \nolimits_{w \in \W} \barPrw \feat ~~~~~~ \forall k\\ 
\end{array}
\label{pro:max_ent}
\end{equation}


In the event a prior distribution about $\PW$ is given, we may use the Principle of Minimum Cross Entropy~\citep{shore1980axiomatic} to replace the above objective with $min ~D_{KL}(\PW \mathrel{\Vert} \Pi(W))~=~$ $\sum\nolimits_{w \in \W} \Prw \log { \frac{\Prw}{\Pi(w)}}$, where $D_{KL}$ is the Kullback-Leibler divergence and $\Pi(W)$ is the prior.  By setting $\Pi(W)$ to the uniform distribution we arrive at the Principle of Maximum Entropy again.  The solution to this program is the well known Gibbs distribution, 
\begin{align}
\lambdaphiPrw = \frac{\Pi(w)e^{\sum \nolimits_k \lambda_k \feat}} {Z_\phi(\lambda)}
\label{eq:gibbsdistr}
\end{align} where $\lambda$ are the parameters of the Lagrangian dual of program~\ref{pro:max_ent} and $Z_\phi$ is the partition function.  The distribution estimation task is thus reduced to choosing $\lambda$ such that the one maximum entropy distribution which matches the given feature expectations $\closure{\Phi}$ is found.


\subsection{Empirical Relaxation}
\label{sec:empapprox}



In many scenarios $\closure{\Phi}$ is not precisely known and instead samples of $w$ from the unknown distribution are available with which we may estimate $\closure{\Phi}$.  When the number of samples is limited we may not be able to apply the Principle of Maximum Entropy as $\closure{\Phi}$ may be invalid, with no choice of $\lambdaphiPW$ able to produce an equal $\Phi$.  Let $\truePW$ be the \enquote{true} distribution from which $N$ samples are pulled i.i.d. to produce $\empPW$.  In certain circumstances we are able to relax the feature expectation constraint in program~\ref{pro:max_ent} using knowledge of the expected difference between $\truePW$ and $\empPW$.

For instance, suppose $\truePW$ is a normal distribution and $\phi(w) = w$.  Then \cite{skilling_maximum_1984} show the following chi-squared relaxation for the feature expectation constraint.

\begin{center}$\chi^2 ~=~  (\Phi - \tilde{\Phi})^2 / \sigma^2$\end{center}

Where $\tilde{\Phi}$ is the sample mean and $\sigma^2$ the sample standard deviation.  This produces an ellipsoid centered around $\empPW$ within which all $\PW$ are equally acceptable given the available data.  Because the sample mean is a consistent estimator this holds as  $N \rightarrow \infty$ with the ellipsoid converging to a point at $\truePW$.  To generalize the above relaxation to multivariate normal distributions we may use the Mahalanobis Distance $(\Phi - \tilde{\Phi})^\top\Sigma^{-1}(\Phi - \tilde{\Phi})$ where $\Sigma$ is the covariance matrix of $\empPW$.

Given the maximum entropy objective function, we have two possible solution modes.  If the uniform distribution is within the ellipsoid we take this maximum possible entropy solution as the answer and interpret this as the provided data was insufficient to extract any information.  Otherwise, the answer will lie somewhere on the ellipsoid boundary itself, making the relaxation an equality constraint.

\subsection{Communications Channel Model}



Other relaxation choices exist, Poisson~\citep{skilling_maximum_1984} for example. In this work, however, we consider a broader set of scenarios where the above relaxation can not apply, for instance where the received data is of lower dimensionality than $\W$ due to missing information from hidden variables or corruption by noise for which a consistent estimator does not exist. Towards the goal of accurately modeling the process of parameter estimation in these cases, we first describe a simple discrete memoryless communications channel.

\begin{figure}[h]
	\centering
	\includegraphics[width=0.6\textwidth]{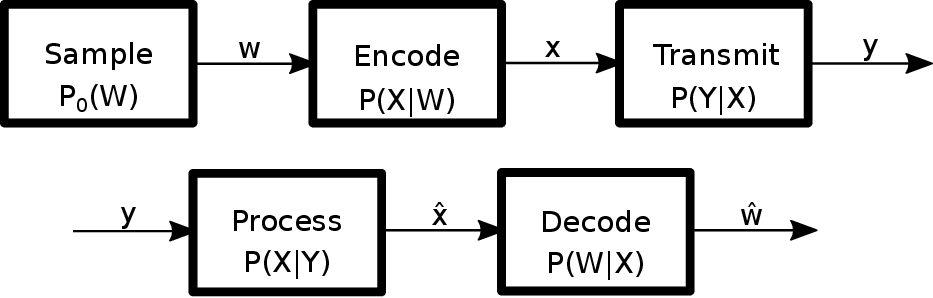}
	\caption{A discrete, memoryless communications channel showing the process of transmitting a message $w$ (top) and receiving (bottom)}
	\label{fig:comm_model}
	\vspace{-0.1in}
\end{figure}

Suppose a \emph{message} $w$ is to be sent to a receiver.  We model each $w$ as being sampled i.i.d. from a distribution $\truePW$, which is unknown to the receiver, with the distribution of all sent messages $\empPW$.  To transmit the message, it must first be physically encoded in the world using some method.  In the communications domain this method is commonly a function of $w$ to aid in decoding by the receiver, here we model this method using the distribution $\PXW$ and assume this distribution is known to the receiver.  This change is to allow for the possibility of messages from non-engineered or partially engineered sources in which $w$ only partially determines the \emph{encoding} $x$.

After encoding, $x$ is sent over some channel to the receiver, where it is received as the \emph{signal} $y$ after possibly being transformed by the channel.  This process is modeled with the distribution $\PYX$, which is known by the receiver.  In the domain of communications, the receiver desires an estimate of the message, $\hat{w}$, which is determined by first processing $y$ to estimate $\hat{x}$ using $\PXY$, and then decoding $\hat{x}$ into $\hat{w}$ using $\PWX$. By contrast in distribution estimation the receiver desires an estimate of the parameters of $\truePW$ using the distribution of all received signals $\empPY$.

A foundational result in information theory~\citep{shannon} shows that it is possible to choose an encoding function such that the decoding error is vanishingly small.  As an example, suppose we have a discrete $\W$, an encoding that maps all $w$ to equally spaced points along the real number line, and a channel that adds zero-mean Gaussian noise to $x$.  We can therefore choose the distance between each $x$ such that it is large compared to the standard deviation of the channel noise. Thus, a large enough distance will ensure the estimate $\hat{x}$ for each $y$ has arbitrarily small error, and thus we may also estimate $\empPW$ and then $\tilde{\Phi}$ with an arbitrary small amount of error. 


\section{Uncertain Feature Expectations}
\label{sec:empPW}


Unfortunately, in many scenarios in which we wish to apply the Principle of Maximum Entropy it is not possible to choose an encoding such that there is arbitrarily low estimation error per $w$ and therefore $\tilde{\Phi}$ may remain uncertain.  To apply the principle or a relaxation like in section~\ref{sec:empapprox} we must choose a single $\Phi$ among those that match available information.  Toward this approach we precisely define the feasible $\PW$ focusing entirely on the communicated information.  Given an arbitrarily accurate estimate of $\PY$, we have:

\begin{align}
\sum \nolimits_{w \in \W} \Prw \sum \nolimits_{x \in \X} \Prxw \Pryx = \Pry ~~ \forall y
\label{eqn:empPY}
\end{align}


We may employ linear algebra to determine all satisfying $\PW$ as follows. Gather all $\Prw$ into the vector $\mathbb{W}$, $\Pry$ into the vector $\mathbb{Y}$, and define the $|\Y|x|\W|$ matrix $A$ as $A_{y,w} = \sum \nolimits_{x \in \X} \Prxw \Pryx$.  Now: $A\mathbb{W}^T = \mathbb{Y}^T$.  We can readily see that the null space of matrix A (and any one satisfying $\PW$) defines the set of all $\PW$ that can produce $\PY$ using the above equation.  When the null space is non-empty, such as when $|\Y| < |\W|$, we have a linear set of $\PW$ any of which equally can produce $\PY$, call this set $O_{\PYW}(\PY)$.  Notice $\truePW$ must be in this set.

In this situation existing methods could be used to choose one $\PW$ so that we may then apply the Principle of Maximum Entropy.  For example, we could employ Baye's law and choose the most likely $\PW$ given a prior on $\truePW$ by solving:

\begin{align}
\Prw = \sum\limits_{y} \Pry ~\delta (w, \argmax_{w'} P(W=w'|Y=y))
\label{eqn:ml}
\end{align}


Where $\delta$ is the Kronecker delta.  However, while a simple solution, we are not guaranteed that the found $\PW$ will satisfy equation~\ref{eqn:empPY} and this would ensure $\PW$ could not be $\truePW$, therefore the corresponding feature expectations could only be correct by chance. Instead, we could employ the Principle of Maximum Entropy to arrive at the $\PW$ with the maximum entropy that satisfies equation~\ref{eqn:empPY}.

\begin{equation}
\begin{array}{l}
 \max -\sum\nolimits_{w \in \W} \Prw \log \Prw \\
  \mbox{{\bf subject to}}\\
  \sum \nolimits_{w \in \W} \Prw = 1 \\
  \sum \nolimits_{w \in \W} \Prw \sum \nolimits_{x \in \X} \Prxw \Pryx = \Pry ~~ \forall y
\end{array}
\label{pro:max_ent_emppw}
\end{equation}


However, without feature constraints we do not incorporate the known structural information into our choice. If we nevertheless solve this program and again apply the Principle of Maximum Entropy by solving program~\ref{pro:max_ent}, setting $\barPW$ to the result of program~\ref{pro:max_ent_emppw}, there is no guarantee that the final found $\PW$ which satisfies equation~\ref{eqn:featureexpectations} will also satisfy equation~\ref{eqn:empPY}, \uline{except in the special case that $\PW = \barPW$}. With this insight, we further develop this approach towards ensuring that all constraints are satisfied.

\section{Uncertain Maximum Entropy}
\label{sec:uMaxEnt}

Define the function $M_\phi \colon \barPW \rightarrow \lambdaphiPW$ as the optimal solution to the $\barPW$ parameterized program~\ref{pro:max_ent}.  As program~\ref{pro:max_ent} is convex the set $\dotphiPW$ of all solutions must be a convex set~\citep{boyd2002convex} containing all distributions that satisfy equation~\ref{eqn:featureexpectations} each having the maximum entropy of all other $\PW$ with equal $\Phi$.  All distributions that match the structural information about $\truePW$ are located somewhere in $\dotphiPW$. Now, if given the set $O_{\PYW}(\PY)$ containing all distributions satisfying information received through a communications channel, we \textbf{must} have $\truePW \in \{\dotphiPW ~\cap~ O_{\PYW}(\PY) \}$.

Let us take an example.  Suppose for $|\W| = 3$, $|\Y| = 2$ we have the following functions: \\

\begin{tabular}{ccc}
\begin{minipage}{0.45\textwidth}
    \begin{flushleft}
      $P(Y|W) = $
      \begin{tabular}{r|c|c|c}
      & $w_1$ & $w_2$ & $w_3$ \\
      \hline
      $y_1$ & 0.95&	0.55	&0.95\\
      $y_2$ & 0.05&	0.45	&0.05
      \end{tabular}
    \end{flushleft}
  \end{minipage} &

  \begin{minipage}{0.27\textwidth}
    $\PY$ = (0.9, 0.1)
  \end{minipage}

  \begin{minipage}{0.27\textwidth}
    $\phi_1(W)$ = (1, 0, 0)
  \end{minipage}

\end{tabular}\\

\begin{figure}[h]
	\centering
  \includegraphics[width=2in]{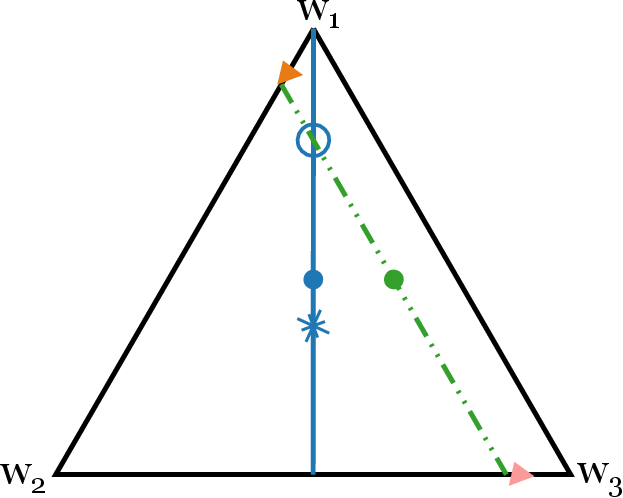}
	\caption{\small $\PW$ probability simplex for the system described above and possible solutions: program~\ref{pro:max_ent_emppw} (green solid dot), applying program~\ref{pro:max_ent} to the results of program~\ref{pro:max_ent_emppw} (solid blue dot), two possible solutions using equation~\ref{eqn:ml} and a uniform prior (light triangle) or using the true distribution as the prior (darker triangle), and the distribution at the intersection of both sets (unfilled blue circle). For reference, the distribution with the maximum possible entropy is marked with a star, solid blue vertical line is $\dotphiPW$, and the dashed green line is $O_{\PYW}(\PY)$. }
	\label{fig:triangle}
	\vspace{-0.1in}
\end{figure}

Given this single feature function $\dotphiPW$ contains the distributions that give equal probability mass to $w_2$ and $w_3$.  We illustrate this as the vertical blue line in the probability simplex in figure~\ref{fig:triangle}. All $\PW$ that satisfy equation~\ref{eqn:empPY} are shown as the dashed green line.  Notice the point at the intersection of the two lines marked with an unfilled circle is the only distribution that satisfies both program~\ref{pro:max_ent} and equation~\ref{eqn:empPY}, and therefore must be $\truePW$.  The feature expectations for this distribution may then be computed, and program~\ref{pro:max_ent} solved to recover $\lambda$.

While in this example there is only a single satisfying solution, in general it can be the case that the system is under-determined and therefore more than one equally satisfying distribution exists.  As we still desire the best estimate of $\lambda$ we must arrive at a single $\Phi$, which corresponds in the worst case to picking a single distribution within this feasible set.  To this end we propose again applying the Principle of Maximum Entropy to choose the one satisfying distribution with the maximum entropy.  Intuitively we define this new \textbf{Principle of Uncertain Maximum Entropy} as when choosing among multiple equally satisfying distributions that match both known structural information and information received through a communications channel, the best choice is the one with the maximum entropy.

This new principle presented as a convex non-linear program with feature expectation and discrete memoryless communication channel constraints is given by:

\begingroup
\renewcommand{\arraystretch}{1.2}
\begin{equation}
\begin{array}{l}
  \max\limits_{\PW} -\sum\nolimits_{w \in \W} \Prw \log \Prw \\
  \mbox{{\bf subject to}}\\
  \sum\nolimits_{w \in \W} \Prw = 1  \\
  \PW \in \{\dotphiPW \cap O_{\PYW}(\PY) \} 
\end{array}
  \label{pro:u_max_ent_program}
\end{equation}
\endgroup


\section{Analysis and Interpretation}

With the goal of justifying our claims in sections~\ref{sec:empPW} and~\ref{sec:uMaxEnt}, we begin our analysis by showing that the Principle of Uncertain Maximum Entropy as shown in program~\ref{pro:u_max_ent_program} is convex.

\begin{kdb_theorem}
\label{thm:umaxentsolutionconvex}
Program~\ref{pro:u_max_ent_program} is convex.
\end{kdb_theorem}

\begin{proof} 
Equation~\ref{eqn:empPY} is linear in $\PW$ and $\dotphiPW$, being the solution set of a convex program, is a convex set. As the intersection of a convex set and a linear set must be convex, the satisfying set is convex.  The objective function is Shannon entropy, which is known to be convex. As a convex program is defined as a convex objective function constrained with a convex feasible set, program~\ref{pro:u_max_ent_program} is a convex program.
\end{proof}

\begin{kdb_corollary}
\label{thm:umaxentunique}
The solution to the principle of uncertain maximum entropy is unique.
\end{kdb_corollary}

Since in a convex program any local optima must be the global optimum the Principle of Uncertain Maximum Entropy may be thought of as a function $U_{\phi, \PYW} \colon \PY \to  \lambdaphiPW$.

\begin{kdb_corollary}
\label{thm:umaxenthighestentropy}
No other satisfying $\PW$ exists with more entropy than the unique solution.
\end{kdb_corollary}

Given that entropy is the opposite of information, we conclude that any other satisfying choice contains more information than the one with maximum entropy.  Given that a higher entropy equally satisfying alternative exists, the added information from any other choice can not be justifiably attributed to the constraints, which represent all knowledge of $\truePW$.  Therefore, this added information must be a form of bias.  All else being equal, the solution with the least bias is expected to be least wrong most often. These facts allow us to arrive at the conclusion that justifies the principle; any other satisfying choice is expected to be worse averaged over all scenarios than that with the maximum entropy $\PW$.

The Principle of Uncertain Maximum Entropy generalizes the classic principle and the Principle of Latent Maximum Entropy (see Appendix~\ref{app:latentmaxent}).


\begin{kdb_theorem}
\label{thm:stdmaxent}
The Principle of Maximum Entropy is a special case of the Principle of Uncertain Maximum Entropy in which only one $\PW$ satisfies equation \ref{eqn:empPY}. 
\end{kdb_theorem}

\begin{proof}
In the event that only one $\PW$ is possible given $\PY$ then it must be $\truePW$, which must be within $\dotphiPW$. We may reduce program~\ref{pro:u_max_ent_program} to only $\PW = \lambdaphiPW$ as $\dotphiPW \cap \{ \truePW \} = \lambdaphiPW$ and the objective and other constraint are trivially satisfied.  As $\lambdaphiPW = M_\phi(\truePW)$ this is identical to program~\ref{pro:max_ent}.
\end{proof}

As $\truePW$ must be located within the feasible set and the Principle of Uncertain Maximum Entropy dictates we choose the member of this set with the maximum entropy, we must have that $H(\truePW) \leq H(U_{\phi, \PYW}(\PY))$.  Therefore, some amount of information may be inaccessible to the receiver so we must consider this information lost to the \enquote{environment}, either through the encoding or transmission, when such a communication channel is in use. Let $E_\phi(W; Y) = H(U_{\phi, \PYW}(\PY)) - H(\truePW)$ be the amount of information lost in this way.  Without knowing $H(\truePW)$ we cannot determine how much information is lost, however, we may upper bound $E_\phi(W; Y)$ by finding the minimal entropy of any $P(W)$ within the satisfying set and then $E_\phi(W; Y) \leq H(U_{\phi, \PYW}(\PY)) - \min H$.  We may now see that the Principle of Uncertain Maximum Entropy produces an upper bound on $H(\truePW)$, or equivalently, a lower bound on its information content.

\section{Multichannel Communications}

The Principle of Uncertain Maximum Entropy readily generalizes to multichannel scenarios.  Suppose a given message's encoding is received simultaneously over more than one channel by the receiver.  We simply consider the joint signal distribution as: $\Y = \Y_1 \times \Y_2$ and alter $\PYX$ accordingly, for instance if $\Y_1$ is independent of $\Y_2$ then $\PYX = P(Y_1|X)P(Y_2|X)$.

Alternatively, if separate messages sampled  from the same $\truePW$ are sent over unique channels, we simply restrict the feasible set further for each unique encoding and channel in use. Let $\mathbb{C}$ be the set of these communication channel tuples $c$: ($\X, \Y, \PXW, \PYX, \PY$). Now for each $c \in \mathbb{C}$ we define a set $O^c$ of $\PW$ which satisfy equation~\ref{eqn:empPY} for $c$.  Then the constraint in program~\ref{pro:u_max_ent_program} becomes:

\begin{equation}
\begin{array}{l}
  \PW \in \{\dotphiPW \bigcap_{c \in \mathbb{C}} O^c \} 
\end{array}
\label{eq:u_max_ent_multi_channelprogram}
\end{equation}

\section{Solution}
\label{sec:alg}


While program~\ref{pro:u_max_ent_program} is convex this assumes that the set $\dotphiPW$ is exactly expressed in closed form.  This is rarely the case in practice as this set is all solutions of an optimization problem.  To solve this program then, we begin by transforming it into a bilevel optimization program.  The Principle of Uncertain Maximum Entropy expressed as an optimistic CP-CP bilevel program is:

\begingroup
\renewcommand{\arraystretch}{1.2}
\begin{equation}
\begin{array}[t]{l}
  \max\limits_{\PW \in \Delta, \barPW} -\sum\nolimits_{w \in \W} \Prw \log \Prw \\
  \mbox{{\bf subject to}}\\
  \sum\nolimits_{w \in \W} \Prw = 1  \\
  \sum \nolimits_{w \in \W} \Prw \sum \nolimits_{x \in \X} \Prxw \Pryx = \Pry ~~~~~~~~~~~ \forall y \\
  \barPW \in \begin{array}[t]{l}
 \max\limits_{\barPW \in \Delta} -\sum\nolimits_{w \in \W} \barPrw \log \barPrw \\
  \mbox{{\bf subject to}}\\
  \sum \nolimits_{w \in \W} \barPrw = 1 \\
  \sum \nolimits_{w \in \W} \barPrw \feat  = \sum \nolimits_{w \in \W} \Prw \feat ~~~ \forall k\\ 
  \end{array} \\
  \PW = \barPW
\end{array}
\label{pro:u_max_ent_bilevel_program}
\end{equation}
\endgroup

Where $\Delta$ is the set of all distributions over $W$.  Notice the coupling constraint $\PW = \barPW$.  This constraint is critical to connect the upper and lower programs, without it the two programs are separately solvable and produce the same answer as solving program~\ref{pro:max_ent_emppw} followed by program~\ref{pro:max_ent}.\footnote{We call this procedure \algname{dMaxEnt} for \enquote{double} MaxEnt}

Solving bilevel programs can present a challenge as in general they are NP-Hard~\citep{jeroslow1985polynomial}.  Even \enquote{simple} linear-linear programs may have a non-convex and disconnected feasible set. Modern approaches include the Branch-and-Bound~\citep{moore1990mixed} and Branch-and-Cut~\citep{tahernejad2020branch} algorithms. As our lower level program is convex we use a technique first developed by~\citep{fortuny1981representation} in which the lower program is replaced by its Karush-Kuhn-Tucker (KKT) conditions, which are necessary and sufficient in this case, to produce the following single level program.:

\begingroup
\renewcommand{\arraystretch}{1.2}
\begin{equation}
\begin{array}[t]{l}
  \max\limits_{\PW, \barPW, \lambda, \eta} -\sum\nolimits_{w \in \W} \Prw \log \Prw \\
  \mbox{{\bf subject to}}\\
  \sum\nolimits_{w \in \W} \Prw = 1  \\
  \sum \nolimits_{w \in \W} \Prw \sum \nolimits_{x \in \X} \Prxw \Pryx = \Pry ~~ \forall y \\
  -\log \barPrw + \eta + \sum\nolimits_{k} \lambda_k \feat = 1 ~~~~~~~~~~~~~~~~ \forall w\\
  \sum \nolimits_{w \in \W} \barPrw = 1 \\
  \sum \nolimits_{w \in \W} \barPrw \feat  = \sum \nolimits_{w \in \W} \Prw \feat ~~~~~~~ \forall k\\ 
  \PW = \barPW
\end{array}
  \label{pro:u_max_ent_bilevel_program_kkt}
\end{equation}
\endgroup

Where $\eta$ and $\lambda$ are the Lagrangian multipliers for the lower program's normalization and feature expectation constraints respectively. Simplifying by replacing $\barPW$ with $\PW$, we arrive at:

\begingroup
\renewcommand{\arraystretch}{1.2}
\begin{equation}
\begin{array}{l}
 \max\limits_{\PW, \lambda, \eta} -\sum\nolimits_{w \in \W} \Prw \log \Prw \\
  \mbox{{\bf subject to}}\\
  \sum \nolimits_{w \in \W} \Prw = 1 \\
  \sum \nolimits_{w \in \W} \Prw \sum \nolimits_{x \in \X} \Prxw \Pryx = \Pry ~~ \forall y \\
  -\log \Prw + \eta + \sum\nolimits_{k} \lambda_k \feat = 1  ~~~~~~~~~~~~~~~~ \forall w
\end{array}
\label{pro:u_max_ent_bilevel_simplified}
\end{equation}
\endgroup

Notice this is a simple modification to program~\ref{pro:max_ent_emppw} that adds a constraint set for the known structural information to the existing constraints for normalization and communicated information.  
\section{Experimental Performance}

We experimentally compare the performance of the Principle of Uncertain Maximum Entropy to a number of existing techniques in randomly generated programs.  We sample from $\truePW$ $N$ times (where $N$ may be infinite), then with each $w$ we sample $x \sim  P(X|W=w)$ and $y \sim  P(Y|X=x)$ respectively, these samples are used to generate $\empPW$ and $\empPY$ as indicated. The \algname{L-BFGS}~\citep{LBFGS} algorithm is used to solve the Lagrangian dual of all programs except \algname{uMaxEnt}. The algorithms we examine are:

\begin{itemize}
\item \algname{uMaxEnt} - Program~\ref{pro:u_max_ent_bilevel_simplified} is solved using $\empPY$ and Sequential Least-Squares Quadratic Programming (see Appendix~\ref{app:experiment}).
\item \algname{MaxEnt} - Program~\ref{pro:max_ent} is solved using $\empPW$, providing a best-case control.
\item \algname{mlMaxEnt} - $\PW$ is computed using the most likely $w$ from equation~\ref{eqn:ml} for each $y$ sample, then program~\ref{pro:max_ent} is solved.
\item \algname{dMaxEnt} - $\PW$ is computed by solving program~\ref{pro:max_ent_emppw} given $\empPY$, then program~\ref{pro:max_ent}  is solved.
\end{itemize}

Where indicated we consider the Kullback-Lieblier divergence $D_{KL} (\lambdaphiPW  \mathrel{\Vert} \truePW )$ as our metric of accuracy.  We use $|\W| = |\X| = 10$ for all experiments. More details are available in Appendix~\ref{app:experiment}.

We first note for $N = \infty$ that all algorithms are equivalent when $W = X = Y$, and in our tests all achieve an average $D_{KL}$ slightly below the stopping tolerance as expected.  In the event that $|\Y| = |\W|$ but $Y \neq W$,  \algname{mlMaxEnt} dramatically reduces in performance relative to all other algorithms as can be seen in table~\ref{tab:y10performance}. \algname{uMaxEnt} and \algname{dMaxEnt} both produce low average error but \algname{uMaxEnt} is significantly better.

\begin{table}[h]
\centering
\begin{tabular}{l|l|l|l|l}
& \algname{MaxEnt} & \algname{uMaxEnt} & \algname{dMaxEnt} & \algname{mlMaxEnt} \\
$Y = W$ &  \num[scientific-notation=true]{0.0000000000000004} &	\num[scientific-notation=true]{0.0000000000000006} &	\num[scientific-notation=true]{0.0000000000000009}	&	\num[scientific-notation=true]{0.0000000000000004} \\
$|\Y| = |\W|$ & \num[scientific-notation=true]{0.0000000000000006} &	\num[scientific-notation=true]{0.00000092883} &	\num[scientific-notation=false]{0.035221} &	\num[scientific-notation=true]{1.13107}\\
\end{tabular}
\caption{Average $D_{KL}$ of the experimentally tested algorithms when $|\Y| = |\W|$, $|\phi|$ varies from 1 to 9.}
\label{tab:y10performance}
\vspace{-0.1in}
\end{table}

\begin{figure}[ht]
	\centering
	\includegraphics[width=0.70\textwidth]{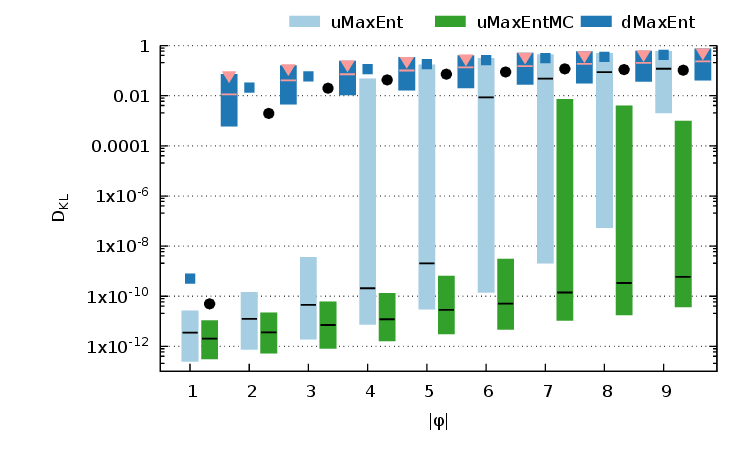}

	\caption{Box plot of the $D_{KL}$ achieved by \algname{uMaxEnt}, \algname{dMaxEnt}, and Multi-Channel \algname{uMaxEnt} for two channels as $|\phi|$ is varied.  Points mark the respective mean, median is the line.}
	\label{fig:performance_feature_size}
	\vspace{-0.25in}
\end{figure}

\begin{figure}[ht]
	\centering
    \includegraphics[width=0.70\textwidth]{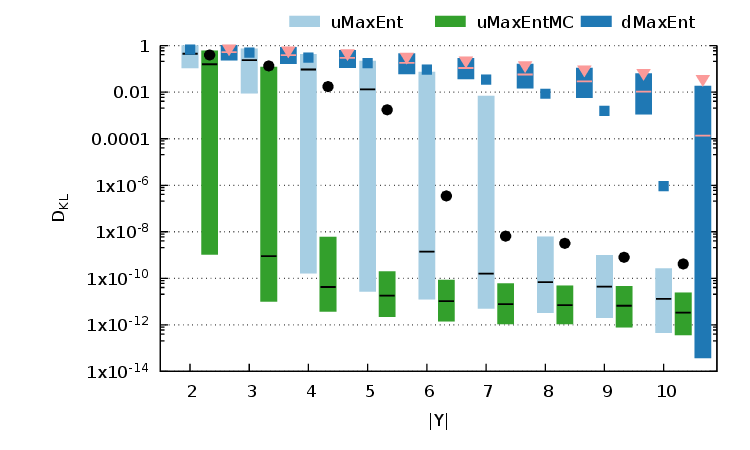}

	\caption{Box plot of the $D_{KL}$ achieved by \algname{uMaxEnt}, \algname{dMaxEnt}, and Multi-Channel \algname{uMaxEnt} for two channels as $|\Y|$ is varied.  Points mark the respective mean, median is the line. }
	\label{fig:performance_y_size}
	\vspace{-0.1in}
\end{figure}

We observe two prominent modes in our \algname{uMaxEnt} results, where over all experimental configurations except $Y = W$, about \num{57}\% of runs achieve a $D_{KL} \leq $ \num[scientific-notation=true]{0.0000001} with the remaining \num{43}\% producing higher error. The primary factors determining the performance of \algname{uMaxEnt} are the number of features $|\phi|$ and the size of the signal alphabet $|\Y|$. As can be seen in figure~\ref{fig:performance_feature_size}, the smaller the number of features the lower the average $D_{KL}$ achieved, with \algname{uMaxEnt} always lower than \algname{dMaxEnt}. Importantly, the interquartile range in figure~\ref{fig:performance_feature_size} shows \algname{uMaxEnt} dramatically outperforms \algname{dMaxEnt} when features are below 6, with median $D_{KL}$ values many orders of magnitude smaller.  This is attributed to the more tightly constrained feasible set producing a smaller area for $\truePW$ to be in, which \algname{uMaxEnt} exploits, finding solutions with less entropy than \algname{dMaxEnt}.  By contrast, \algname{dMaxEnt} is not as significantly affected by the number of features, with both the average and median $D_{KL}$ reducing by about one order of magnitude as the features go from 9 to 1.

\begin{figure}[h]
    \centering
	\includegraphics[width=0.70\textwidth]{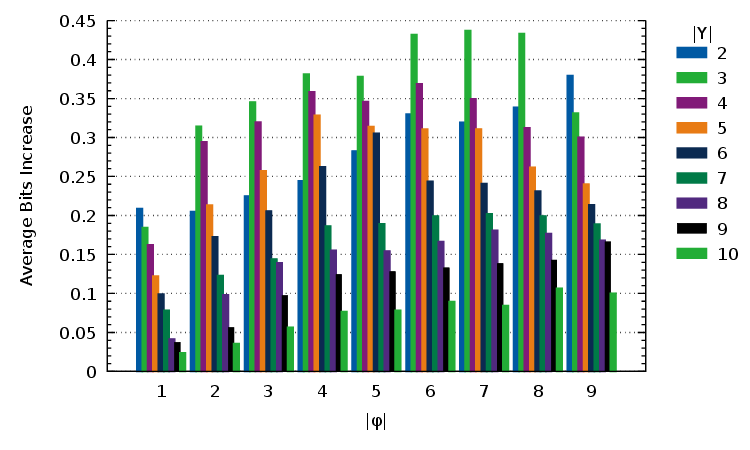}
	\caption{Information increase in the solutions found by \algname{uMaxEnt} relative to that found by program~\ref{pro:max_ent_emppw} alone for varying sizes of $|\phi|$ and $|\Y|$, $Y \neq W$.}
	\label{fig:additional_info}
	\vspace{-0.1in}
\end{figure}

Because \algname{uMaxEnt} is producing solutions with a minimal amount of bias, we may attribute a reduction in entropy relative to the feature-free program~\ref{pro:max_ent_emppw} to the information contained in the features.  However, as we show in figure~\ref{fig:additional_info} where we compare \algname{uMaxEnt} against program~\ref{pro:max_ent_emppw}, we cannot consider the information gain from features separate from $Y$.  Notice that the information increase is usually highest when $|\Y|$ is low and therefore provides less information about $\truePW$ than when $|\Y|$ is higher. Note that less entropy in this case does not necessarily mean a corresponding increase in accuracy.  We show table~\ref{tab:exact_perform} in appendix~\ref{app:exact_stat_sig} indicating which combinations produce statistically significant better accuracy compared to \algname{dMaxEnt}.

Figure~\ref{fig:performance_y_size} examines the performance of our algorithms as $|\Y|$ is varied.  Below $|\Y| = 6$, most solutions are in the high error mode of our results and above 6 in the low error mode.  Overall, the smaller $|\Y|$ is the less information is available in a given communications channel, and we see an expected increase in error in our results.

\subsection{Multi-Channel \algname{uMaxEnt}}

\begin{figure}[ht]
	\centering
    \includegraphics[width=0.70\textwidth]{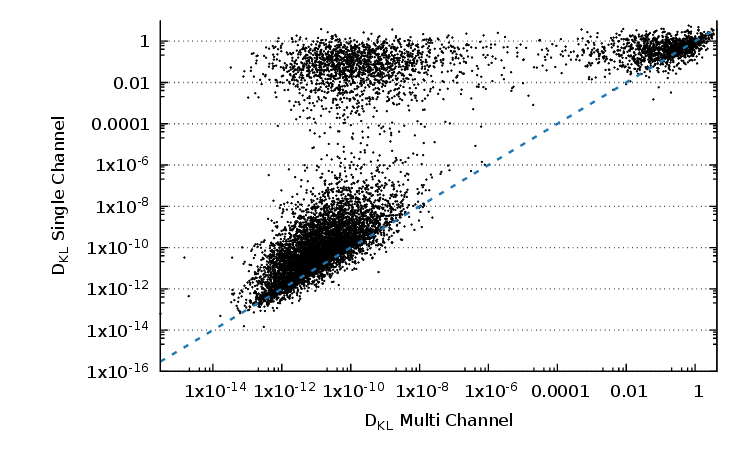}

	\caption{Scatter plot of the $D_{KL}$ achieved by multichannel \algname{uMaxEnt} against single-channel \algname{uMaxEnt}, 20,000 points randomly selected from our data set (where $Y \neq W$). Note log scales on X and Y axes.}
	\label{fig:performance_multi}
	\vspace{-0.1in}
\end{figure}

Due to the loss of information when $\Y \neq \W$, we cannot expect \algname{uMaxEnt} to produce low error solutions for all possible communication channels.  However, by adding additional channels the feasible set can be constrained further leading to lower error.  We show this in figures~\ref{fig:performance_feature_size} and \ref{fig:performance_y_size} as the multichannel \algname{uMaxEntMC} algorithm with just one additional channel significantly outperforms single-channel \algname{uMaxEnt} for all feature counts and $|\Y|$.  We visualize this increase in performance further with a scatter plot in figure~\ref{fig:performance_multi}, notice particularly the top-left cluster; these points have a high $D_{KL}$ found by single-channel \algname{uMaxEnt} but with just one extra channel the $D_{KL}$ has dropped dramatically.  The top-right cluster shows that some high $D_{KL}$ results are still present and whose error may be improved upon with additional channels.

\subsection{uMaxEnt Relaxation for Finite Samples}
\label{sec:approx}


\algname{uMaxEnt} requires that $\truePY$ is estimated to arbitrary accuracy so that $\truePW$ will lie within $O_{\PYW}(\PY)$. However, this is only guaranteed in scenarios where $\truePY$ is perfectly known or the number of available samples is infinite.  For all other scenarios we must relax constraints appropriately.

\begin{itemize}
  \item \algname{uMaxEnt}: In our experiments using discrete $Y$, we model the samples of $y$ as being pulled i.i.d. from a Multinomial distribution and use its conjugate prior, the Dirichlet Distribution, to define a distribution over all $\PY$ given the samples.  We then approximate the Dirichlet using a multivariate Gaussian distribution and relax the communication constraint of program~\ref{pro:u_max_ent_bilevel_simplified} to use a similar $\chi^2$ relaxation to that described in section~\ref{sec:empapprox}, except in $Y$ instead of $W$.  In our experiments, we use a p-value of 0.05.
  
  There are two possible solution types to consider, if the $\PY$ computed from a uniform $\PW$ lies within the covariance ellipsoid centered at $\empPY$, then this solution is chosen.  Otherwise, due to the linear relationship between $\PW$ and $\PY$, $\PW$ is chosen such that the computed $\PY$ lies on the covariance ellipsoid and $\PW$ has the maximum entropy of all other choices that have a $\PY$ on or within the ellipsoid. 
  
  As the covariance matrix of a Dirichlet distribution is singular, we cannot use it to compute the Mahalanobis distance.  To remedy this, we remove the last row and column of the covariance matrix, ignore the last entry of the vectors representing distributions over $Y$, and use this reduced Mahalanobis distance with a $|\Y| - 1$ degree of freedom $\chi^2$ metric.

  \item \algname{dMaxEnt}: For $N < \infty$, $\empPY$ may be poorly estimated and all satisfying $\PW$ may be outside the probability simplex.  To address this, we assume a Gaussian prior on $\PY$ and add a Gaussian regularization term $ \sum\nolimits_k \lambda_k^2 / N $ to the Lagrangian dual of program~\ref{pro:max_ent_emppw}.  For small $N$, we expect similar performance to \algname{uMaxEnt} as the $\PW$ chosen is forced to be very close to uniform.
\end{itemize}

\subsubsection{Relaxation Performance}

\begin{figure}[h]
	\centering
	\includegraphics[width=0.75\textwidth]{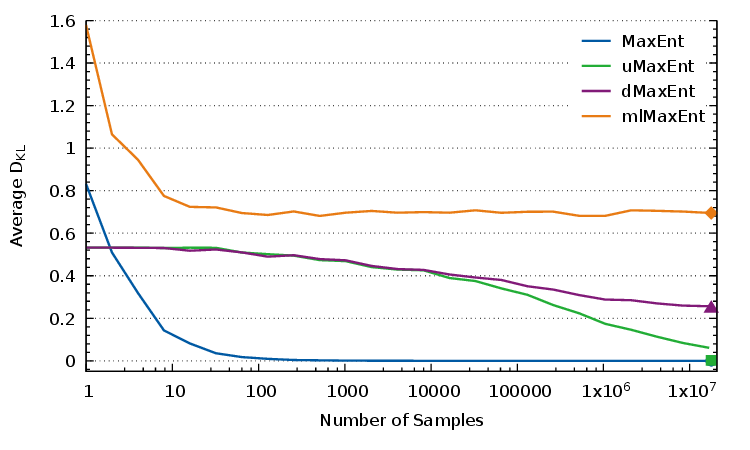}
	\caption{Average $D_{KL}$ achieved by varying algorithms as the number of samples increases.  Performance at $N = \infty$ is marked with a point. Feasible solutions only, $|\Y| = 4, \alpha = 3, |\phi| = 2$. Note log scale on the X axis.}
	\label{fig:d_vs_u}
	\vspace{-0.1in}
\end{figure}

We show the typical performance of our relaxation in figure~\ref{fig:d_vs_u} for $|\Y| = 4, \alpha = 3, |\phi| = 2$.  The performance of non-regularized \algname{MaxEnt} and \algname{mlMaxEnt} are shown for reference, in particular the low performance of \algname{mlMaxEnt} demonstrates the extreme challenge of this problem domain.  Notice the points showing performance at infinity, indicating that \algname{uMaxEnt} will still significantly improve with further samples while \algname{dMaxEnt} is near its maximum performance at $N = 10^7$. 

The observed  typical relative performance of \algname{uMaxEnt} and \algname{dMaxEnt} is described by three phases.  In the first phase, when $N$ is very small ( $\approx <$ 12), both algorithms choose nearly uniform $\PW$ owing to their Gaussian approximation / regularization.   Phase 2 is marked by a lack of statistically significant difference between the two algorithms. As $N$ increases both gradually improve their performance by approximately the same amount, in this case by about $20\%$.  In the third phase \algname{uMaxEnt} outperforms \algname{uMaxEnt} as $N$ increases towards $\infty$.    

Given this behavior we may ask: how many samples are needed to reach phase 3?  Given table~\ref{tab:exact_perform} in appendix~\ref{app:exact_stat_sig} showing which combinations of $|\Y|$ and $|\phi|$ achieve statistically significant better performance at infinity, we answer this question with table~\ref{tab:approx_perform} in appendix~\ref{app:exact_stat_sig} showing the minimum samples (as powers of 2) needed for our \algname{uMaxEnt} relaxation to achieve statistically significant better performance than \algname{dMaxEnt}.  This performance is influenced by the choice of p-value; higher p-values reduce the size of the ellipsoid leading to potentially lower error solutions at the cost of increasing the risk of no feasible solution existing and vice versa.  As a specific example, in table~\ref{tab:pvalue} we show the average $D_{KL}$ and feasible proportion of experiment runs for all $N$ sizes in a single configuration as the p-value is varied, clearly showing the tradeoff and the sensitivity of our algorithm to the $\chi^2$ ellipsoid parameters.

\vspace{-0.1in}
\begin{table}[h]
  \centering
\begin{tabular}{ l | c c c }
 & \multicolumn{3}{c}{p-value}\\
  & 0.1 & 0.05 & 0.01 \\
 \hline 
$D_{KL}(\PW \mathrel{\Vert} \truePW )$ & \num{1.90537} & \num{1.99501} & \num{2.77611} \\
\% Feasible & 96.560\% & 97.619\% & 99.206\% \\
\end{tabular}
\caption{\algname{uMaxEnt} Average error and proportion of executions that found a feasible result as the p-value is varied for $|\Y| = 5, \alpha = 3, |\phi| = 3$ across $N$ from 1 - 1,048,576 }
\label{tab:pvalue}
\vspace{-0.5in}
\end{table}

\section{Discussion}

In our experiments the use of one additional communication channel showed a reduction in the average $D_{KL}$ and therefore less information loss on average.  Ignoring the special case where $W = Y$, as we continue adding random channels to $\mathbb{C}$ we expect that the set of $\truePW$ for which more than one $\PW$ satisfies the multichannel constraints goes to zero as $|\mathbb{C}| \rightarrow \infty$.  Therefore, we may interpret the classic Principle of Maximum Entropy as an infinite-channel Principle of Uncertain Maximum Entropy, where there is no possible information lost to the environment and the entropy is minimal with no way of increasing information further without adding bias. 

In this work we assume that both $\phi$ and $\PYW$ are known to the receiver.  Future work includes relaxing these assumptions, for instance, if $\phi$ is not completely known $\truePW$ is expected to have unmodeled structural variables or otherwise unmodeled noise. In this situation, \algname{uMaxEnt} must be modified as $\truePW$ may not lie within $\dotphiPW$.  For instance, suppose we have some bound on the maximum divergence between any point in $\dotphiPW$ and $\truePW$, call it $\epsilon_M$.  The coupling constraint in program~\ref{pro:u_max_ent_bilevel_program} could be replaced with $D_{KL}(\PW \mathrel{\Vert} \barPW) \leq \epsilon_M$. This will ensure we find the point consistent with observations and the maximum allowed divergence from $\dotphiPW$.

Improved performance of relaxed \algname{uMaxEnt} was shown for large but finite $N$, however, no significant increase for smaller $N$ over \algname{dMaxEnt} with Gaussian prior was observed.  As our algorithm is sensitive to the acceptance region, here approximated as an ellipsoid despite the potentially asymmetric structure of Dirichlet distributions, a more accurate approximation may improve performance for smaller $N$ significantly.

\acks{This work was partially funded by NSF grant 1830421.  The authors wish to thank our colleagues in the computer science department at UNCA for their support
and comments.}

\vskip 0.2in
\bibliography{bibliography}

\appendix

\section{}
\label{app:latentmaxent}

The Principle of Latent Maximum Entropy as presented by~\cite{wang2012} is a generalization of the Principle of Maximum Entropy in which some portion of each message variable $w$ is missing to the receiver and the rest is received without error.  Divide each $w$ into $y$ and $z$ such that $w = y \cup z$ and $\PW = P(Y, Z)$.  $y$ is the portion of $w$ that is perfectly received while $z$ is perfectly hidden, and let $\Z_y$ be the set of all $z$ which when combined with a given $y$ forms a $w \in \W$.  The principle of latent maximum entropy is now defined as:

\begin{align}
&  \max \left( -\sum\nolimits_{w \in \W} \Prw \log \Prw \right )\nonumber\\
&  \mbox{{\bf subject to}} \nonumber\\
& \sum \nolimits_{w \in \W} \Prw = 1 \nonumber\\
&  \sum \nolimits_{w \in \W} \Prw \phi_k(w)  = \sum \nolimits_{y \in \Y} \Pry \sum\nolimits_{z \in \Z_y} Pr(z | y) \phi_k(w)  ~~~~~~ \forall k 
\label{pro:latent_max_ent}
\end{align}

The authors give an Expectation-Maximization~\citep{mclachlan2007algorithm} based algorithm, \algname{EM-IS}, to find solutions to program~\ref{pro:latent_max_ent}.  The relevant steps of this algorithm are:

\begin{itemize}
\item E Step: Compute $\closure{\Phi} = \sum \nolimits_{y \in \Y} \Pry \sum\nolimits_{z \in \Z_y} Pr(z | y) \phi_k(w)~~ \forall k$
\item M Step: Solve program~\ref{pro:max_ent}
\end{itemize}

\begin{kdb_theorem}
\label{thm:latentmaxent}
The Principle of Latent Maximum Entropy as solved with the expectation-maximization based \algname{EM-IS} algorithm with an initial $\Prw = 1/|\W|$ is a special case of the Principle of Uncertain Maximum Entropy in which $|\Y| < |\W|$ and $\PYW$ is deterministic. 
\end{kdb_theorem}

\begin{proof}
As $Pr(y, z) = \Prw$ we may rewrite the constraint of program~\ref{pro:latent_max_ent} as:
\begin{align}
\sum \nolimits_{w \in \W} \Prw \phi_k(w)  = \sum \nolimits_{y \in \Y} \frac{\Pry}{\barPry} \sum\nolimits_{z \in \Z_y} Pr(w) \phi_k(w)  \nonumber
\end{align}

Where $\barPry = \sum\nolimits_{z \in \Z_y} \Prw$.  Let $Pr(y|w) = 1$ if $\exists z \in \Z_y \colon w = (y \cup z)$, otherwise $0$.  This constraint is satisfied whenever $\barPry = \Pry = \sum\nolimits_{w \in \W} \Prw P(y|w) ~~\forall y$. Therefore, program~\ref{pro:latent_max_ent} as given is equivalent to program~\ref{pro:max_ent_emppw}.

Now notice that the \algname{EM-IS} algorithm restricts the search space of $\PW$ to $\dotphiPW$ by only returning distributions of the form $\lambdaphiPW$ from the M step.  As \algname{EM} monotonically increases towards convergence, when the \algname{EM} algorithm converges the found $\lambdaphiPW$ will satisfy the constraints of program~\ref{pro:u_max_ent_program}.  Because the search space is convex it will not converge early to a local optima.  Beginning the search at the maximum entropy $\PW$ possible completes the proof.
\end{proof}

\section{}
\label{app:experiment}

\textbf{Additional Experimental Details}\\

To directly compare the performance of the four algorithms considered, we give each one identical input, or as close to identical as possible.  For instance, if $\truePY$ is given to \algname{uMaxEnt}, we give the corresponding $\truePW$ used to generate it to \algname{MaxEnt}. We describe the steps to accomplish this below.

\subsection{Model Generation}

We first fix $|\W| = 10$ and select $|\phi| \in \{ 1, 2, \dots, 9 \}$.  We generate sparse feature sets by choosing a probability $p \in \{0.25, 0.33, 0.66\}$ and then assigning $\phi_k(w) = 1$ with probability $p$ for each $k, w$, otherwise 0.  If a given $\phi_k = 0 ~\forall ~w$, we re-generate this feature until at least one $w$ is assigned a $1$.  We choose true weights $\lambda_k = U(-1, 1)* \alpha$, where $\alpha$ is 2 or 3, now each $\truePrw \propto e^{\sum\limits_k \lambda_k \feat }$.   We then sample $\truePW$ \num{16777216} times to provide the samples needed for the relaxed experiments. 

\subsection{Channel Generation}

Given the above model $\truePW$ and samples, choose values $\alpha_x \in \{ DET, 0.5, 1, 2\}$ and $\alpha_y \in \{ DET, 0.5, 1, 2\}$.  If $\alpha_x = DET$, set $X = W$ and if $\alpha_y = DET$, set $Y = X$. Otherwise, fix $|X|$ at 10 and then choose $|Y| \in \{ 2, 3, \dots, 10\}$. Now let $Pr(x|w) \propto e^{\alpha_x U(0, 1)} ~\forall~ w$, $Pr(y|x) \propto e^{\alpha_y U(0, 1)} ~\forall~ x$ respectively.  We use a block structure in our experiments where all permutations of these parameters are equally represented in our data set.  

Using these distributions and $\truePW$, we calculate $\truePX$ and $\truePY$ which is used in the  (infinite $N$) experiments.  For the relaxed experiments, we generate \num{16777216} samples of $X$ and $Y$ by sampling $x_i \sim  P(X|w_i)$ and $y_i \sim  P(Y|x_i)$ respectively.

This process is repeated a second time for the multichannel experiments, except we hold $|X|$ and $|Y|$ the same, and sample $\alpha_x$ and $\alpha_y$ from $U(0, 2)$.  We examine only exact multichannel results.

\subsection{Procedure}

Experiments proceeded first by loading a given model and communication channel(s), then generating $\empPW$ and $\empPY$ using the first $N$ samples of each respectively, or setting $\empPW = \truePW$ and $\empPY = \truePY$ for exact experiments.  Each algorithm is then input its appropriate distribution ($\empPW$ for \algname{MaxEnt}, $\empPY$ all others) and solved.  The resulting $\lambdaphiPW$'s were recorded and $D_{KL}(\lambdaphiPW \mathrel{\Vert} \truePW)$ was calculated.

We use the NLOpt~\citep{NLopt} library version v2.10.0 to solve our programs, specifically the Lagrangian Dual of \algname{MaxEnt}, \algname{mlMaxEnt}, and \algname{dMaxEnt} programs are solved with \algname{L-BFGS}~\citep{LBFGS}, while \algname{uMaxEnt} is solved with \algname{SLSQP}~\citep{SLSQP}.  Tolerances were set as low as possible to ensure accuracy while avoiding numerical instability, with change in parameter tolerance \num[scientific-notation=true]{0.000000000000001}, maximum evaluations \num{1000000}, and absolute maximum parameter value of \num{1000}.

As program~\ref{pro:u_max_ent_bilevel_simplified} is non-convex and \algname{SLSQP} is a feasible-path algorithm we are not guaranteed to find the global optimum parameters without feasible initial starting conditions.  However, as the dimensionality of the problem increases it may be difficult or impossible to find a satisfying initial point in all cases.  Our approach is to choose small random values for $\lambda_k \sim U(-0.1, 0.1)$, compute $\PW$ using equation~\ref{eq:gibbsdistr}, and set $\eta = -\log(\sum\limits_w e^{\sum\limits_k \lambda_k \feat}) + 1$. This ensures at least the structural and normalization constraints are initially satisfied.

Additionally, implementation and numerical stability issues may cause early convergence or failure during line search causing the algorithm to terminate without a feasible solution even if one exists.  In practical terms, this means the outcome of \algname{uMaxEnt} is partially dependent upon the choice of initial starting conditions.  We incompletely address these concerns by repeating the above process 10 times, choosing new initial conditions each time, and of the feasible solutions found we take the one with the highest entropy.  In the event no feasible solution is found and $N = \infty$ we increase tolerance and repeat as a feasible solution must exist.  If $N < \infty$ we simply mark the problem as infeasible.

\section{}
\label{app:exact_stat_sig}

\textbf{uMaxEnt Performance Tables}\\
\vspace{-2em}
\begin{table}[h]
  \centering
\begin{tabular}{ c | c c c c c c c c c }
\multicolumn{10}{ c }{$|\Y|$} \\
$|\phi|$ & 2 & 3 & 4 & 5 & 6 & 7 & 8 & 9 & 10 \\
\hline
1 & \textbf{0.2277} & \textbf{0.1863} & \textbf{0.1580} & \textbf{0.1036} & \textbf{0.0575} & \textbf{0.0423} & \textbf{0.0147} & \textbf{0.0178} & \textbf{0.0099} \\
2 & \textbf{0.2006} & \textbf{0.2869} & \textbf{0.2695} & \textbf{0.1863} & \textbf{0.1225} & \textbf{0.0770} & \textbf{0.0609} & \textbf{0.0268} & \textbf{0.0103} \\
3 & \textbf{0.0922} & \textbf{0.3338} & \textbf{0.3101} & \textbf{0.2244} & \textbf{0.1657} & \textbf{0.0894} & \textbf{0.0738} & \textbf{0.0675} & \textbf{0.0301} \\
4 & \textbf{0.1225} & \textbf{0.1470} & \textbf{0.3155} & \textbf{0.2898} & \textbf{0.2143} & \textbf{0.1410} & \textbf{0.0933} & \textbf{0.0623} & \textbf{0.0385} \\
5 & 0.0245 & \textbf{0.1449} & \textbf{0.2169} & \textbf{0.3280} & \textbf{0.2680} & \textbf{0.1715} & \textbf{0.0921} & \textbf{0.0837} & \textbf{0.0407} \\
6 & 0.0452 & 0.0184 & \textbf{0.1622} & \textbf{0.2183} & \textbf{0.1877} & \textbf{0.1666} & \textbf{0.1334} & \textbf{0.0958} & \textbf{0.0607} \\
7 & 0.0121 & 0.0318 & 0.1029 & \textbf{0.1584} & \textbf{0.1943} & \textbf{0.1665} & \textbf{0.1490} & \textbf{0.1092} & \textbf{0.0604} \\
8 & 0.0778 & \textbf{0.2072} & \textbf{0.1019} & \textbf{0.1414} & \textbf{0.1621} & \textbf{0.1896} & \textbf{0.1277} & \textbf{0.1190} & \textbf{0.0900} \\
9 & 0.0559 & \textbf{0.0850} & \textbf{0.2389} & \textbf{0.1700} & \textbf{0.0992} & \textbf{0.1789} & \textbf{0.1652} & \textbf{0.1609} & \textbf{0.0705} \\
\end{tabular}
\caption{Table comparing the accuracy of program~\ref{pro:max_ent_emppw} with program~\ref{pro:u_max_ent_bilevel_simplified}, which differs from \ref{pro:max_ent_emppw} only in the additional structural constraint, by showing the average\\
 $\left ( D_{KL}( \PW^\eqref{pro:max_ent_emppw} \mathrel{\Vert}  \truePW) - D_{KL}(\PW^\eqref{pro:u_max_ent_bilevel_simplified} \mathrel{\Vert}  \truePW) \right )$ on problems from our experiments organized by the number of features and $|Y|$.  Both programs were given the same problems to solve. Statistically significant averages (p $<$ 0.05, paired t-test, right tailed) marked in bold, more than 100 problems generated per data point, $\alpha = 3$.
}
\label{tab:exact_perform}
\end{table}

\begin{table}[h]
  \centering
\begin{tabular}{ c | r r r r r r r r r }
  \multicolumn{10}{ c }{$|\Y|$} \\
$|\phi|$ & 2 & 3 & 4 & 5 & 6 & 7 & 8 & 9 & 10 \\
\hline 
1 & 12 & 18 & 17 & 19 & 20 & 22 & 20 & 23 & 23 \\
2 & 16 & 15 & 16 & 16 & 15 & 20 & 21 & 21 & 23 \\
3 & 17 & 18 & 15 & 16 & 19 & 20 & 20 & 21 & 22 \\
4 & 22 & 20 & 20 & 17 & 18 & 19 & 21 & 21 & 23 \\
5 &  & 19 & 20 & 24 & 21 & 20 & $>$24 & 22 & $>$24 \\
6 &  &  & 24 & 22 & 22 & 24 & $>$24 & $>$24 & $>$24 \\
7 &  &  &  & $>$24 & $>$24 & $>$24 & $>$24 & $>$24 & $>$24 \\
8 &  & 23 & $>$24 & $>$24 & $>$24 & $>$24 & $>$24 & $>$24 & $>$24 \\
9 &  & $>$24 & $>$24 & $>$24 & $>$24 & $>$24 & $>$24 & $>$24 & $>$24 \\
\end{tabular}
\caption{Approximately the number of samples needed (as powers of 2) for relaxed \algname{uMaxEnt} to produce statistically significant better performance on average than \algname{dMaxEnt}. Paired t-test, right-tailed, p = 0.05, only feasible solutions considered. }
\label{tab:approx_perform}
\end{table}

\end{document}